\documentclass[10pt,journal,compsoc]{IEEEtran}
\usepackage[style=ieee]{biblatex}
\usepackage{beramono}
\usepackage{caption} 
\usepackage[inline]{enumitem}
\usepackage[hidelinks]{hyperref}
\usepackage{listings}
\usepackage{multirow}
\usepackage{siunitx}
\usepackage{tabularx}
\usepackage[colorinlistoftodos,prependcaption]{todonotes}
\usepackage{tikz}
\usepackage{xspace}
\usepackage{xcolor}
\usepackage{float}
\usepackage{mdframed}

\setuptodonotes{inline}

\newcommand*{\eg}{e.g.\@\xspace}
\newcommand*{\ie}{i.e.\@\xspace}

\lstdefinelanguage{diff}{
basicstyle=\ttfamily\small,
morecomment=[f][\color{diffstart}]{@@},
morecomment=[f][\color{diffincl}]{+},
morecomment=[f][\color{diffrem}]{-},
}

\definecolor{diffinclColor}{RGB}{50,220,30}
\definecolor{diffremColor}{RGB}{230,50,50}
\definecolor{diffstart}{named}{gray}
\definecolor{diffincl}{named}{diffinclColor}
\definecolor{diffrem}{named}{diffremColor}

\makeatletter
\newcommand*{\etc}{%
    \@ifnextchar{.}%
        {etc}%
        {etc.\@\xspace}%
}
\makeatother

\newcommand{\toolname}{\textsc{Supersonic}\xspace}

\makeatletter
\newenvironment{btHighlight}[1][]
{\begingroup\tikzset{bt@Highlight@par/.style={#1}}\begin{lrbox}{\@tempboxa}}
{\end{lrbox}\bt@HL@box[bt@Highlight@par]{\@tempboxa}\endgroup}

\newcommand\btHL[1][]{%
  \begin{btHighlight}[#1]\bgroup\aftergroup\bt@HL@endenv%
}
\def\bt@HL@endenv{%
  \end{btHighlight}%
  \egroup
}
\newcommand{\bt@HL@box}[2][]{%
  \tikz[#1]{%
    \pgfpathrectangle{\pgfpoint{1pt}{0pt}}{\pgfpoint{\wd #2}{\ht #2}}%
    \pgfusepath{use as bounding box}%
    \node[anchor=base west, fill=yellow!30,outer sep=0pt,inner xsep=1pt, inner ysep=0pt, rounded corners=0pt, minimum height=\ht\strutbox+1pt,#1]{\raisebox{1pt}{\strut}\strut\usebox{#2}};
  }%
}
\makeatother
\lstdefinestyle{Highlight}{
    moredelim=**[is][\btHL]{`}{`},
    moredelim=**[is][{\btHL[fill=orange!50]}]{´}{´},
    moredelim=**[is][{\btHL[fill=red!50]}]{@}{@},
}

\mdfsetup{nobreak=true}

\bibliography{references.bib}

\begin{document}

\title{\toolname: Learning to Generate \\Source Code Optimizations in C/C++}

\author{Zimin~Chen,
        ~Sen~Fang,
        and~Martin~Monperrus%
\IEEEcompsocitemizethanks{\IEEEcompsocthanksitem Zimin Chen, Sen Fang and Martin Monperrus are with KTH Royal Institute of Technology, Sweden.\protect\\
E-mail: \{zimin, senf, monperrus\}@kth.se}%
}

\IEEEtitleabstractindextext{%
\begin{abstract}
Software optimization refines programs for resource efficiency while preserving functionality. Traditionally, it is a process done by developers and compilers. This paper introduces a third option, automated optimization at the source code level. We present \toolname, a neural approach targeting minor source code modifications for optimization. Using a seq2seq model, \toolname is trained on C/C++ program pairs ($x_{t}$, $x_{t+1}$), where $x_{t+1}$ is an optimized version of $x_{t}$, and outputs a diff.
\toolname's performance is benchmarked against OpenAI's GPT-3.5-Turbo and GPT-4 on competitive programming tasks.
The experiments show that \toolname not only outperforms both models on the code optimization task but also minimizes the extent of the change with a model more than 600x smaller than GPT-3.5-Turbo and 3700x smaller than GPT-4. %
\end{abstract}

}

\maketitle

\IEEEdisplaynontitleabstractindextext

\IEEEpeerreviewmaketitle

\section{Introduction}
\label{sec:intro}
\IEEEPARstart{S}{oftware} optimization refers to the process of refining a program so that it utilizes fewer resources, such as time, memory, CPU, and energy while preserving its original functionality. Traditionally, this task has been carried out by the developer and/or the compiler. The developer possesses deep human expertise to enhance the program by using a more efficient data structure or an algorithm with better complexity. On the other hand, the compiler can apply a range of automated optimizations on the intermediate representation that can significantly enhance the program's performance. Human developers optimize at the source code level, and compilers at the machine code level. In this work, we explore a third way: automated optimization at the source code level.

For developers, automatic source code optimization is invaluable. Firstly, it encompasses optimizations that are beyond the scope of compiler optimizations, high-level optimizations that a compiler cannot achieve with guarantees. For instance, refactoring an inefficient algorithm or modifying data structures to boost performance is beyond the compiler's scope. Secondly, it facilitates the optimization of legacy systems for which language and domain expertise have been lost over time, such as Fortran libraries or Cobol systems. Process-wise, automatic source code optimization can be used in modern code bases, with automated pull requests \cite{sadowski2015tricorder}.

In this paper, we introduce \toolname, an innovative end-to-end system that employs a supervised machine-learning approach to the problem of automatic source code optimization. \toolname, implemented as a seq2seq model, learns the relationship between input programs and their optimized versions. Our process involves collecting a dataset of past source code optimizations, where each pair ($x_{t}$, $x_{t+1}$) consists of a base program $x_{t}$ and its optimized counterpart $x_{t+1}$. Special attention has been given to creating such a high-quality training dataset. With this dataset, we then tailor a model and devise a data and training loop specifically for the optimization task. 

At inference time, \toolname has three phases: canonicalization, diff-synthesis, and post-processing. During the canonicalization phase, \toolname removes source code comments and ensures a canonical code format (w.r.t. whitespaces, tabs, \etc) that has also been enforced at training time. The diff-synthesis phase takes the input program and predicts a diff-based output representation, which is similar to the typical Unix patch format. The post-processing phase is responsible for validating the predicted diff-based output representation and applying it to the original program. \toolname is implemented for C/C++ and trained on a large corpus of C/C++ program optimizations.

For the training and evaluation of \toolname, we use programs from three code competition websites. This is a perfect evaluation benchmark since problems on code competition websites are usually designed to have a large solution space and they also have a strong emphasis on optimization, running time and memory consumption. They are excellent for collecting training data as well as for assessing optimized programs. For evaluation, we compare against a strong baseline. \toolname competes with OpenAI's GPT-3.5-Turbo and GPT-4, two industrial large language models that have proven to be state-of-the-art in many software engineering tasks~\cite{tian2023chatgpt,liu2023improving,sobania2023analysis,poldrack2023ai}. Our evaluation of 559 programs shows that \toolname outperforms both GPT-3.5-Turbo and GPT-4. Not only is the performance better on the task of optimizing C/C++ programs, but \toolname's model is approximately 600x smaller\footnote{We assume it is the same size as GPT3, which is a 175B parameter model.} than GPT-3.5-Turbo and 3700x smaller\footnote{There are no official numbers on the number of parameters of GPT-4, we assume it is a trillion parameter model.} than GPT-4, making it more cost and energy-efficient. 

\toolname is unique compared to the most related work, such as PIE4Perf~\cite{madaan2023learning} and DeepDev-PERF~\cite{garg2022deepdev}.
First, we take special care in removing essentially full re-implementations, and not optimized solutions from the dataset. Second, \toolname uses a novel diff-based output representation as opposed to generating the entire program, a feature that, as demonstrated in our evaluation, significantly boosts its effectiveness. Last, the evaluation of \toolname is done on third-party competition websites, giving our results strong external validity.
    
In summary, our contributions are:
\begin{itemize}
\item \toolname, a novel source code optimization technique based on state-of-the-art sequence-to-sequence learning. \toolname is able to generate source code level optimizations while retaining significant similarity with the original program.
\item We show that \toolname outperforms GPT-3.5-Turbo and GPT-4 when submitting tentative optimizations to the official Codeforces website. It improves running time for 26.0\% programs, compared to only 12.0\% for GPT-3.5-Turbo and 4.0\% for GPT-4.
\item We investigate the optimization performance of \toolname, GPT-3.5-Turbo, and GPT-4 at various string similarity thresholds between the optimized and original program. We find that \toolname is better than GPT-3.5-Turbo and GPT-4 when the threshold is above 0.4. However, the threshold of 0.4 is close to a complete rewrite.
\item We demonstrate that the diff-based output representation is better than outputting the full program. Our ablation study shows that changing the full program output representation to a diff-based output representation of \toolname boosts the optimization success rate by 2x at least.
\item For the sake of open science and future research on this topic, we share all our code, data, and train models at \url{https://github.com/ASSERT-KTH/Supersonic}.
\end{itemize}

\section{Background}
\label{sec:background}

\subsection{Large language models}
\label{sec:background:language_model}
A Large Language Model (LLM) is a type of deep learning (DL) model commonly used in the field of natural language processing (NLP). Unlike traditional DL models designed for specific tasks, LLMs are initially trained on large textual datasets to acquire a universal language representation. This learned representation can be further refined and adapted for various downstream tasks through supervised fine-tuning \cite{devlin2018bert}. The majority of current LLMs are built upon the core modules of the Transformer model \cite{vaswani2017attention}, and they can be categorized into three main types based on the chosen modules: encoder-only \cite{devlin2018bert}, decoder-only \cite{radford2018improving}, and encoder-decoder \cite{lewis2019bart}.

The design of an encoder-only LLM involves stacking multiple Transformer encoder layers, with BERT \cite{devlin2018bert} emerging as a renowned representative of this LLM class. BERT specifically undergoes training with two training objectives - masked language modeling and next sentence prediction. The former lets BERT predict masked words according to their located context, and the latter lets BERT measure whether two sentences are continuous, both of which allow it to learn a universal contextual representation of words. In contrast, decoder-only LLMs are structured around the Transformer decoder as their fundamental building block. The GPT family \cite{radford2018improving, brown2020language}, as one of the most well-known decoder-only LLMs, is pre-trained with an objective called next token prediction. Here, the models are tasked to predict the present word based on its preceding context. The encoder-decoder LLM, as the name suggests, incorporates both Transformer encoder and decoder layers. T5 \cite{raffel2020exploring}, a prototypical example of this LLM category, is pre-trained with an objective analogous to a fill-in-the-blank task, which compels the model to predict missing words in relation to their specific context.

Due to the inherent differences in LLMs' core architectures and training objectives, encoder-only and decoder-only LLMs respectively excel in language understanding and generation tasks. On the other hand, encoder-decoder LLMs, while capable of delivering high performance across both types of tasks, necessitate more parameters for their construction.

LLMs are increasingly utilized in the field of software engineering (SE) to expedite automation processes. An example of this is CodeBERT \cite{feng2020codebert}, an LLM designed specifically for programming language. This model was initially trained on a broad, cross-programming language (PL) corpus. Subsequently, supervised fine-tuning allows its application to various SE tasks, such as code searching \cite{gu2018deep} and summarization \cite{hu2018deep}. To augment the efficacy of LLMs in handling code, \citeauthor*{wang2021codet5} introduced CodeT5 \cite{wang2021codet5}, which was pre-trained using an innovative method known as identifier-aware denoising training. Additional training objectives include identifier tagging, masked identifier prediction, and bimodal dual generation. All of these objectives facilitated the model's ability to effectively comprehend both programming and natural language. In recent developments, next-generation LLMs like LLaMa \cite{touvron2023llama}, BLOOM \cite{scao2022bloom}, and PaLM \cite{chowdhery2022palm} are gaining momentum. These models are typically trained on both extensive PL and natural language corpora. As a result, they offer robust support for tasks within both NLP and SE domains.

\begin{figure*}[t]
    \centering
    \includegraphics[width=1\linewidth]{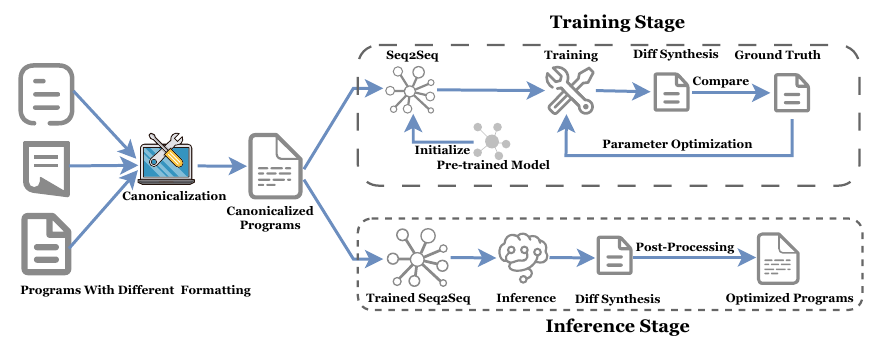}
    \caption{The pipeline of \toolname for code optimization.}
    \label{fig:pipeline}
\end{figure*}

\subsection{The State-of-the-art Models at OpenAI}

Recently, OpenAI released GPT-3.5-Turbo \cite{openaichatgpt} and GPT-4 \cite{openai2023gpt4}, two groundbreaking NLP systems specializing in generating text that closely mirrors human language. These state-of-the-art systems owe their performance to the Transformer decoder-based architecture, known as GPT \cite{radford2018improving}. GPT-3.5-Turbo's training involves a two-stage process: unsupervised pre-training and subsequent instruction fine-tuning. In the pre-training phase, the model learns from an extensive corpus of Internet data dated until September 2021. It is important to underscore that the system does not possess specific knowledge about the documents that comprise its training data and cannot directly access any database or document. The second stage, instruction fine-tuning, involves training the model on a smaller dataset with human feedback, utilizing a technology called reinforcement learning from human feedback \cite{ouyang2022training}. With suitable instructions, also called prompts, GPT-3.5-Turbo can serve a wide array of applications, including but not limited to generating code~\cite{liu2023improving}, debugging~\cite{surameery2023use}, acting as a tutoring assistant\cite{tian2023chatgpt}, translating languages~\cite{jiao2023chatgpt}, and even simulating characters for video games~\cite{park2023generative}. Following the success of GPT-3.5-Turbo, OpenAI introduced GPT-4, an even more advanced model with an increased number of parameters and refined training techniques. The GPT-4 is better than GPT-3.5-Turbo on understanding and generating text, improving its predecessor in terms of context accuracy, diversity of output, and adaptability across diverse applications~\cite{openai2023gpt4}. In this work, we explore the capability of GPT-3.5-Turbo and GPT-4 in code optimization.

\subsection{Sequence-to-Sequence learning}
\label{sec:background:seq2seq}
Sequence-to-Sequence (Seq2Seq) learning \cite{sutskever2014sequence}, a concept introduced by the NLP community, is a machine learning paradigm that models conditional probability distributions for sequence outputs based on sequence inputs. This is usually implemented through an end-to-end neural network structure: an encoder and a decoder. Specifically, the encoder maps the input sequence into a fixed-length vector. To illustrate, taking code optimization as an example, the encoder could process a section of code requiring optimization, such as a C++ program, and convert it into a compact vector representation. Leveraging the powerful learning capabilities of neural networks, this vector can effectively capture the semantic information of the input sequence. Subsequently, the decoder utilizes this vector to produce the output sequence. Continuing with our example, the decoder could yield a C++ program optimized for some objectives, \eg, execution speed or memory usage. Training Seq2Seq models are typically conducted end-to-end, tuning the parameters to maximize the likelihood of the correct output sequence given the input sequence. Thus, Seq2Seq learning is a robust framework for tackling generation tasks involving intricate, variable-length sequence transformations.

\section{Technical Solution: \toolname}
\label{sec:tool}

\subsection{Overview}
\toolname leverages seq2seq learning initialized with pre-trained models to generate optimized C/C++ programs. Importantly, the optimized programs are syntactically similar to the original program, meaning that the optimization only changes a few lines of the program. It is done by training \toolname on program pairs, ($x_{t}$, $x_{t+1}$), where $x_{t}$ and $x_{t+1}$ are functionally equivalent, with a small edit distance. \toolname consists of three phases:
\begin{enumerate*}
    \item Canonicalization,
    \item Diff synthesis,
    \item Post-processing.
\end{enumerate*}

\autoref{fig:pipeline} shows the training and inference pipeline of \toolname for code optimization from canonicalization to post-processing. The canonicalization phase canonicalizes the program coding style (whitespaces, tabs, and newlines) and removes comments before it is used as the input to the diff synthesis phase. The diff synthesis phase generates multiple optimization candidates in a diff-based output representation. The post-processing phase checks for the well-formedness of the diff-based output representation, and then applies the output as a patch on the original program. For the training stage of \toolname, the file is first canonicalized, then it is used as input to the seq2seq model. The seq2seq model generates a diff-based representation in the diff synthesis phase, and it is compared with the ground truth to update the parameter to the seq2seq model in an iterative loop. For the inference stage of \toolname, the input still goes through the canonicalization and diff synthesis phase, but since there is no ground truth, the output is applied to the original program in the post-processing phase.

In the following sections, we illustrate each phase of \toolname in detail and describe our training dataset.

\subsection{Canonicalize training source code}
\label{sec:tool:pre_processing}

The goal of the canonicalization phase is to canonicalize program pairs ($x_{t}$, $x_{t+1}$) before they are used as input to the machine learning model. In this phase, we first remove source code comments and apply a unique, consistent code style. 

We remove source code comments because they are subjective in nature, varying greatly in style, quality, and content. Meanwhile, although they do not affect the logical flow of the code, including comments makes the input program longer, which impacts negatively on the inference performance of the language model \cite{anil2022exploring}.

Input programs also vary in coding styles, for example number of whitespace, newlines, and how brackets are placed. These are all tokenized and used as input to the machine-learning model. Therefore we use a single coding style to let the machine learning model focus on the functionality of the program. 

We use \textit{GCC} (using the command "\textit{gcc -fpreprocessed -dD -E -P}") to remove source code comments and \textit{clang-format} (using the command "\textit{clang-format --style=llvm}") to format all C/C++ programs according to LLVM coding style.

\subsection{Output representation \& Diff synthesis}
\label{sec:tool:diff_synthesis}

\begin{figure*}[t]
\begin{center}
\begin{minipage}[t]{.3\textwidth}
\begin{lstlisting}[language=C++, columns=flexible, frame=single, basicstyle=\small, label={lst:original_solution}, caption={A user submission to the Green Bin problem at AtCoder}, captionpos=b, breaklines=true, keywordstyle=\color{blue}, commentstyle=\color{dkgreen}, stringstyle=\color{mauve}]
#include <algorithm>
#include <iostream>
#include <map>
using namespace std;
using ll = long long;
int main() {
  ll N, ans{};
  cin >> N;
  map<string, int> m;
  string S;
  while (cin >> S) {
    sort(begin(S), end(S));
    ans += m[S]++;
  }
  cout << ans << endl;
}
\end{lstlisting}
\end{minipage}
\hfill
\begin{minipage}[t]{.3\textwidth}
\begin{lstlisting}[language=Java, columns=flexible, frame=single, basicstyle=\small, label={lst:improved_solution}, caption={The improved solution by the same user. \textit{unordered\_map} that do not sort the mapping is used instead. The changed lines are highlighted in red.}, captionpos=b,breaklines=true, keywordstyle=\color{blue}, commentstyle=\color{dkgreen}, stringstyle=\color{mauve}, style=Highlight]
#include <algorithm>
#include <iostream>
@#include <unordered_map>@
using namespace std;
using ll = long long;
int main() {
  ll N, ans{};
  cin >> N;
@  unordered_map<string, int> m;@
  string S;
  while (cin >> S) {
    sort(begin(S), end(S));
    ans += m[S]++;
  }
  cout << ans << endl;
}
\end{lstlisting}
\end{minipage}
\hfill
\begin{minipage}[t]{.3\textwidth}
\begin{lstlisting}[language=C++,columns=flexible, frame=single, basicstyle=\small,label={lst:diff_output}, caption={The diff-based output representation by \toolname, the `-' and `+' are part of the output of the network.}, captionpos=b, breaklines=true, keywordstyle=\color{blue}, commentstyle=\color{dkgreen}, stringstyle=\color{mauve}]
 #include <iostream>
-#include <map>
+#include <unordered_map>
 using namespace std;

   cin >> N;
-  map<string, int> m;
+  unordered_map<string, int> m;
   string S;
\end{lstlisting}
\end{minipage}
\end{center}
\caption{The comparison between the full program and the diff-based output representation generated by \toolname. When tokenized, the fully optimized program contains 124 tokens whereas the diff-based output representation contains 71 tokens. The difference will be even bigger if the original solution is longer. A longer output increases the probability of generating an erroneous program because of accumulated mistakes.}
\label{fig:full_vs_diff}
\end{figure*}

A key aspect of \toolname is that it generates a diff-based output representation instead of generating the whole optimized program. The diff-based output representation is the Unix unified diff format between $x_{t}$ and $x_{t+1}$ with one line context, \ie one line before and after the change. Compared to the original diff format, the diff header that specifies the filenames and changed line numbers is removed. Multiple change hunks are concatenated with newlines to represent a change that modifies multiple locations. In contrast, most related works \cite{madaan2023learning,garg2022deepdev} employ a full program output representation. We use the diff-based output representation because generating a longer output sequence increases the probability of errors due to accumulated mistakes. It is widely known that seq2seq models performance decreases with the output length \cite{cho2014properties}.

\autoref{fig:full_vs_diff} shows an example of the diff-based output representation generated by \toolname. This example is a user submission to AtCoder - Green Bin problem \footnote{\url{https://atcoder.jp/contests/abc137/tasks/abc137\_c}}. The task is to find the number of anagrams from a list of strings. The original solution, \autoref{lst:original_solution}, uses a sorted mapping \textit{map} to store pairs of strings and integers. The improved solution, \autoref{lst:improved_solution}, by the same user used \textit{unordered\_map} that does not sort the mapping compared to \textit{map}. This change improved the running time from 144 ms to 97 ms, and memory consumption from 11136 B to 10660 B. \toolname's diff-based output representation of \autoref{lst:improved_solution} is shown in \autoref{lst:diff_output}. The diff-based output representation contains 71 tokens instead of 124 tokens when tokenizing the full improved program. As we will see in the evaluation, the shorter token length helps the model to capture knowledge with fewer tokens.

\subsection{Post-processing synthesized code}
\label{sec:tool:post_processing}

During the inference stage of \toolname, the diff-based output representation from the diff synthesis phase needs to be patched on the original program to get the full source code of the predicted optimized program. We greedily match each predicted changed hunk to the original program using the one-line context, and then apply the changed hunk. If we fail to match the context lines in the original program, it means that the diff synthesis phase has generated a malformed diff-based output representation and we simply discard it.

\subsection{Training loop}
\label{sec:tool:implementation}

The core machine learning model of \toolname that generates and predicts the diff-based output representation is implemented as a transformer-style seq2seq model. The encoder and decoder are initialized with CodeBERT which is pre-trained on C++ code, CodeBERT-CPP \cite{zhou2023codebertscore}. CodeBERT is a pre-trained LLM on source code and natural language~\cite{feng2020codebert}. The training objective of CodeBERT is masked language modeling, where the model predicts the token masked by a special mask token, and replaced token detection, where the model predicts if a token is the original one or not. CodeBERT trained on these two objectives achieves a robust understanding of code semantics and syntax. By initializing our model with weights from CodeBERT, we leverage this rich foundation of code semantics and structure, which accelerates convergence during training and can lead to better overall performance~\cite{sagheer2019unsupervised}.

Our seq2seq model is trained to generate $x_{t+1}$ by using $x_{t}$ as the input of the program pair ($x_{t}, x_{t+1}$) of the training set. We reuse the tokenizer CodeBERT-CPP, which is based on the subword tokenization from WordPiece with \num{50265} as the vocabulary size~\cite{wu2016google}. The total amount of parameters for the seq2seq model is 278M. %

\begin{table*}[]
\begin{center}
\begin{tabular}{llllllll}
\hline
Split      & Size   & $x_{t}$ LOC & $x_{t+1}$ LOC & \toolname output lines & Codeforces solutions & AtCoder solutions & AIZU solutions \\ \hline
Train      & 312876 & (35, 91) & (35, 90)       & (5, 16)              & 276714               & 28665             & 7497           \\
Validation & 1000   & (36, 88) & (36, 88)       & (5, 15)              & 877                  & 96                & 27             \\
Test       & 559  & (33, 86) & (33, 86)       & (5, 17)              & 300                 & -               & 259            \\ \hline
Total      & 314435 & (35, 91) & (35, 90)       & (5, 16)              & 277891               & 28761             & 7783           \\ \hline
\end{tabular}
\end{center}
\caption{Our dataset statistics. The values in parentheses of the $x_{t}$ LOC, $x_{t+1}$ LOC and \toolname output lines columns are the lower and upper quartile.}
\label{tab:dataset}
\end{table*}

\begingroup
\renewcommand{\arraystretch}{1.2} %
\begin{table}[]
\begin{tabularx}{\linewidth}{lX}
\hline
Attribute             & Description                                                \\ \hline
origin                & The origin of the submission: AIZU, AtCoder, or Codeforces \\
author                & The author of the submission                               \\
contest\_id           & ID of the contest/problem                                  \\
submission\_id        & ID of the submission                                       \\
creation\_time        & When the submission is created                             \\
problem               & The problem name                                           \\
programming\_language & The programming language: C or C++                         \\
cpu\_time             & The execution time                                         \\
memory                & The memory consumption                                     \\
source\_code          & The submission source code                                 \\ \hline
\end{tabularx}
\caption{Collected attributes for each submission to AIZU, AtCoder, and Codeforces.}
\label{tab:attributes}
\end{table}
\endgroup

\subsection{Training Dataset}
\label{sec:tool:dataset}
Training \toolname to optimize C/C++ programs requires a dataset of source code optimization program pairs ($x_{t}$, $x_{t+1}$). In our work, we collect source code optimization program pairs from code competition websites. Code competition websites usually record the execution time and memory usage of all submitted solutions to each problem. It makes them the ideal source to crawl high-quality and accurate source code optimization program pairs.

In our work, we use data from Codeforces~\footnote{\url{https://codeforces.com}}, as well as AIZU~\footnote{\url{https://judge.u-aizu.ac.jp}} and AtCoder~\footnote{\url{https://atcoder.jp}} originally from the CodeNet dataset \cite{puri2021codenet} to build our goal dataset. For the Codeforces dataset, we first use the Codeforces API to get all the past contests and submissions to each contest. However, the Codeforces API does not return the source code, therefore we use another source \footnote{https://codeforces.com/blog/entry/94755} to find the source code for each submission. For submissions to AIZU and AtCoder, the CodeNet dataset already provides all the necessary information that we need to process the data. All the submissions that we collect are accepted by each competition website, therefore their correctness is guaranteed. The attributes that we collect from each competition website are shown in \autoref{tab:attributes}.

To ensure the relevance of our dataset, we specifically focused on program pairs ($x_{t}$, $x_{t+1}$) from the same author, where $x_{t+1}$ strictly improves either on the running time or memory usage over $x_{t}$. By requiring the submissions to be in chronological order, we aimed to capture the author's progress and improvements over time. This approach allowed us to observe the evolution of optimizations employed by an author and analyze the specific code changes that led to enhanced performance. Collecting these pairs of programs directly from the same author, in chronological order, minimizes confounding factors such as coding style, enabling us to attribute improved performance to code changes.

The primary objective of \toolname is to generate optimized programs while minimizing the extent of changes from the original code. Intuitively, we want to train the system to generate optimizations that modify a few lines of code only. 
To do this, we filter all pairs of submissions based on the following criteria:
\begin{enumerate}
\item At most 20 lines are changed between $x_{t}$ and $x_{t+1}$.
\item At most 20\% of lines are changed between $x_{t}$ and $x_{t+1}$.
\item At least a string similarity score of 0.8 between $x_{t}$ and $x_{t+1}$.
\end{enumerate}

The string similarity metric provides a value between 0.0 and 1.0, with 0.0 indicating two strings are entirely dissimilar and 1.0 indicating a perfect match between the strings. The string similarity metric we use is the \textit{SequenceMatcher} from the Python \textit{difflib} library. It uses a variant of the Gestalt pattern matching string-matching algorithm \cite{ratcliff1988pattern}.

In total, we collected 20M Codeforces submissions and 6M submissions from the CodeNet dataset. Then, we extracted 746K and 138K submission pairs from Codeforces and CodenNet which either improves on the running time or memory consumption. After filtering the submission pairs, we split the remaining submission pairs into train, validation, and test sets, consisting of \num{312876}, \num{1000}, and \num{559} samples, respectively. The pre-training dataset used to pre-train CodeBERT is the C++ source code of the GitHub-Code-Clean dataset under the CodeParrot project on Huggingface~\footnote{\url{https://huggingface.co/datasets/codeparrot/github-code-clean}}. The full dataset statistics are shown in \autoref{tab:dataset}.

\section{Experimental Setup}
\label{sec:experiment}

In this section, we describe our experimental setup to evaluate \toolname against the state-of-the-art related work.

\subsection{Research Questions}

\begin{enumerate}[align=left]
\item[\textbf{RQ1 - Effectiveness}] How does \toolname compare against GPT-3.5-Turbo and GPT-4 on the task of code optimization while minimizing the extent of changes?
\item[\textbf{RQ2 - Optimization Focus}] How does \toolname compare against GPT-3.5-Turbo and GPT-4 depending on the extent of acceptable changes?
\item[\textbf{RQ3 - Ablation Study}] To what extent is the diff-based output representation better than the full program output representation?
\end{enumerate}

These research questions are framed to address different aspects of the problem domain. RQ1 compares \toolname against GPT-3.5-Turbo and GPT-4 on running time and memory consumption optimizations, while minimizing the extent of changes, by submitting predicted program to code competition websites. RQ2 in contrast to RQ1, relaxes the constraint of minimizing the extent of changes. Lastly, RQ3 explores the impact of using our diff-based output representation compared to using the full program, to validate our core design.

\subsection{Test dataset}
\label{sec:experiment:test_dataset}

Our test dataset, described in \autoref{sec:tool:dataset}, consists of user submissions of code competition problems. The problems are of different difficulties and are designed with possible optimization in mind. These problems are known to be hard for language models. For example, GPT-4 is ranked at the bottom 5\% when trying to Codeforces problems\cite{openai2023gpt4}. The code competition website is also used to compute and report the running time and memory usage.

Our full test dataset consists of 559 user submissions (see \autoref{tab:dataset}), of which 300 are Codeforces user submissions, and 259 are AIZU user submissions. The 559 user submissions, along with their predictions, and the collected running time and memory consumption are used to answer all research questions. Our test set does not overlap with any examples in the pre-training or our own training dataset. The check is done by checking all pairs of examples in the test set and the pre-training dataset for exact matches.

\subsection{Methodology for RQ1}
\label{sec:experiment:rq1}

\begin{table*}
\resizebox{\linewidth}{!}{%
\begin{tabular}{ll|ll|ll|ll}
\multicolumn{2}{l|}{\multirow{2}{*}{Metrics}} & \multicolumn{2}{c}{GPT-3.5-Turbo} & \multicolumn{2}{c}{GPT-4} & \multicolumn{2}{c|}{\toolname} \\
\multicolumn{2}{l|}{}                         & Running time    & Memory & Running time    & Memory    & Running time     & Memory   \\ \hline
Codeforces & \%OPT & 12.0\% (36) & 3.7\% (11) & 4.0\% (12) & 1.3\% (4) & 26.0\% (78) & 8.0\% (24) \\
           & PI    & 3.80$\times$ & 11.57$\times$ & 2.73$\times$ & 1.53$\times$     & 2.65$\times$ & 1.81$\times$ \\ \hline
AIZU       & \%OPT & 4.6\% (12) & 1.9\% (5) & 1.5\% (4) & 0.8\% (2) & 3.5\% (9) & 1.2\% (3) \\
           & PI    & 6.87$\times$ & 3.69$\times$ & 4.88$\times$ & 2.45$\times$ & 2.82$\times$ & 1.23$\times$
\end{tabular}}
\caption{The computed metrics for \num{300} Codeforces user submissions and \num{259} AIZU user submissions for GPT-3.5-Turbo, GPT-4 and \toolname. Within the parentheses for \%OPT is the absolute number of optimized program.}
\label{tab:rq1:codeforces}
\end{table*}

The goal of RQ1 is to study whether \toolname, OpenAI's GPT-3.5-Turbo and GPT-4 can optimize programs. Recall that code optimization typically involves making edits to only a portion of the original program, without completely changing it: \citeauthor{alali2008s} found that the majority of changes are classified as small, changing 6-46 lines~\cite{alali2008s}. With that respect,  the experiment takes care of discarding re-implementations of the original program. Smaller edits are also less likely to contain bugs~\cite{purushothaman2005toward} and more likely to be accepted as a pull request~\cite{weissgerber2008small}.

Concretely, \toolname, GPT-3.5-Turbo and GPT-4 are asked to generate 10 optimization predictions per original program in our test set described in \autoref{sec:tool:dataset}. Then, we filter those where the string similarity value compared to the original program is lower than 0.8, the value we used in \autoref{sec:tool:dataset}, in order to discard re-implementations. Next, the metrics that we use to compare \toolname and GPT-3.5-Turbo are running time and memory usage (from \cite{madaan2023learning}):

\begin{itemize}
    \item \textbf{Percent optimized} - \textsc{\%OPT}: The percentage of optimized programs in the test set.
    \item \textbf{Performance improvement} - \textsc{PI}: The average improvement in running time or memory usage of the best-optimized program among the predictions. If \textit{old} and \textit{new} are running time or memory usage of the original and optimized program, then $\textsc{PI} = \frac{\textit{old}}{\textit{new}}$. We report the average \textsc{PI} over the test set.
\end{itemize}

We compute these metrics for strictly better memory improvements and for strictly better running time improvements. We realize that code competition measurements are not perfectly precise and tend to have some variation. To account for this noise, we only consider a program to be truly more optimized if the performance improvement is at least 20\%, i.e. \textsc{PI} is at least \num{1.2}.

\begin{figure}
\begin{lstlisting}[frame=single, breaklines=true, breakindent=0pt, columns=fullflexible, mathescape=true, caption={The prompt used for querying OpenAI models to generate more optimized programs.}, captionpos=b, label={lst:chatgpt_prompt}]
$\textbf{system prompt:}$ I want you to act as an experienced C and C++ developer and your task is to optimize my written C or C++ programs. I want you to optimize my program for running time and memory usage. I will type my C or C++ program and you will optimize the program and return the optimized program. I want you to only reply with the fixed program inside one unique code block, and nothing else. Do not write explanations.

$\textbf{user prompt:}$ [CODE]
\end{lstlisting}
\end{figure}

We utilize OpenAI's official APIs to call GPT-3.5-Turbo and GPT-4 for generating predictions. As per OpenAI's official API documentation\footnote{https://platform.openai.com/docs/models/gpt-3-5}, the APIs are updated to incorporate the latest model iteration approximately two weeks subsequent to its release\footnote{https://help.openai.com/en/articles/6825453-chatgpt-release-notes}. To ensure clarity regarding the API version employed in our experiments, our predictions spanned from April 17 to July 5 2023, whereas for GPT-4, predictions were conducted between August 21 and August 25 2023. The GPT-3.5-Turbo and GPT-4 prompt we use to generate the predictions is shown in \autoref{lst:chatgpt_prompt}. We use a carefully designed prompt per the best practices\footnote{https://github.com/f/awesome-chatgpt-prompts}. The prompt instructs the model to act as an experienced C/C++ developer and optimize the given program with respect to running time and memory usage. 

\subsection{Methodology for RQ2}
\label{sec:experiment:rq2}

In RQ2, the goal is to study to what extent \toolname, GPT-3.5-Turbo and GPT-4 optimize by re-implementing. To do that, we compare \toolname, GPT-3.5-Turbo and GPT-4 at various string similarity thresholds using the same protocol as RQ1. Also, we study the string similarity distribution for all predictions to see how much the programs are changed for \toolname and GPT-3.5-Turbo. Then, we plot the \textsc{\%OPT} value for different string similarity thresholds between 0 and 1. 

\subsection{Methodology for RQ3}
\label{sec:experiment:rq3}

The goal of RQ3 is to evaluate if the diff-based output representation is a better code representation than the full program representation.
We use the same model architecture and the same dataset split described in \autoref{sec:tool:dataset} and used in RQ1 and RQ2. The only difference is that one model learns to generate the full program, while the other learns to generate the diff-based output representation described in \autoref{sec:tool:diff_synthesis}. Similar to RQ1 described in \autoref{sec:experiment:rq1}, 10 predictions are generated per sample in the test set and predictions with a string similarity value below 0.8 are filtered. The rest are submitted to the corresponding code competition website. We compute \textsc{\%OPT} and \textsc{SP} on running time and memory consumption.

\section{Results}
\label{sec:result}

\subsection{Answer to RQ1 (Effectiveness)}
\label{sec:result:rq1}

\noindent
\begin{figure}
\begin{minipage}{\columnwidth}
\begin{lstlisting}[language=diff, columns=fullflexible, frame=single, basicstyle=\small, keepspaces=true, label={lst:supersonic_optimization_example}, caption={Unified diff for one optimization by \toolname. It simplifies if statements, and removes the need of array \textit{arr}. The string similarity between the original and predicted program is 0.95. The original problem is the \textit{Middle Class} problem on Codeforces\protect\footnotemark.}, captionpos=b, breaklines=true]
@@ -28,12 +28,7 @@
     sort(v.rbegin(), v.rend());
     for (int i = 0; i < v.size(); i++) {
       sm += v[i];
-      if (i == 0)
-        arr[0] = v[0];
-      else {
-        arr[i] = sm / (i + 1);
-      }
-      if (arr[i] >= m)
+      if (sm / (i + 1) >= m)
         cnt++;
     }
     cout << cnt << E;
 #include <algorithm>
 #include <cmath>
\end{lstlisting}
\end{minipage}
\end{figure}
\footnotetext{{\url{https://codeforces.com/problemset/problem/1334/B}}}

\noindent
\begin{figure}
\begin{minipage}{\columnwidth}
\begin{lstlisting}[language=diff, columns=fullflexible, frame=single, basicstyle=\small, keepspaces=true, label={lst:supersonic_compiler_example}, caption={Unified diff for another optimization by \toolname. It add several GCC pragmas that instructs the GCC compiler on how to handle the code. These optimize the code’s execution speed and enable specific compiler optimizations and instruction sets. The string similarity between the original and predicted program is 0.99. The original problem is the \textit{WW} problem on AIZU\protect\footnotemark.}, captionpos=b, breaklines=true]
@@ -1,3 +1,6 @@
+#pragma GCC optimize("Ofast")
+#pragma GCC optimize("unroll-loops")
+#pragma GCC target("avx,avx2,sse,sse2")
 #define _CRT_SECURE_NO_WARNINGS
 #include <algorithm>
 #include <cmath>
\end{lstlisting}
\end{minipage}
\end{figure}
\footnotetext{{\url{https://onlinejudge.u-aizu.ac.jp/problems/1537}}}

\autoref{tab:rq1:codeforces} shows the comparison between \toolname, GPT-3.5-Turbo, and GPT-4 on generating source code optimizations for Codeforces and AIZU submissions. For the Codeforces submissions, \toolname generates optimizations that improve 26.0\% (78) and 8.0\% (24) of the programs on running time and memory consumption respectively. This shows the advantage and efficacy of \toolname over both GPT-3.5-Turbo and GPT-4, illustrating its capability for optimizing a range of problems, even when compared against the most advanced generative models currently available. This is despite the fact that \toolname is 600x and 3700x smaller than GPT-3.5-Turbo and GPT-4.
For AIZU submissions, both \toolname and GPT-3.5-Turbo perform at a similar level for running time and memory optimization, while GPT-4 is still performing worse. The reason why \toolname has a lower performance for AIZU submissions than Codeforces submissions might be because our training data contains more than 80\% program pairs from Codeforces, hence have better learned on that type of problem. The problem set on Codeforces and AIZU is indeed different, Codeforces being competitive in nature with regular contests dividing participants with ratings, AIZU being primarily an educational platform.

\autoref{lst:supersonic_optimization_example} and \autoref{lst:supersonic_compiler_example} are two examples of optimization by \toolname, as a diff between the original and generated program. The first example simplifies the if statements to a single if statement and removes the need for the array \textit{arr} that keeps the intermediate values. The generated example reduces the memory consumption from 2.64MB to 1.56MB, saving 41\% of memory usage. The second example adds three GCC pragma directives that are instructions to the GCC compiler on how to compile the code. \textit{optimize("Ofast")} enables all \textit{-O3} optimizations along with those applicable to all programs compliant with the standard. \textit{optimize("unroll-loops")} unroll loops, a technique that attempts to optimize the running time by sacrificing the binary size. \textit{target("avx,avx2,sse,sse2")} enables the use of specific instruction set extensions for vector operations. The generated program reduces the running time to 7ms from the original 17ms, which is a significant 60\% decrease. This also demonstrates that \toolname is able to instruct the compiler on how to better compile the program, in addition to changing the data structures, compilers, or algorithms.

We give a breakdown of all 5590 predicted programs (10x559 predictions,  for every program in the test set), for \toolname. The post-processing phase of \toolname removed 2045 programs because of malformed diff-based output representation that can not be applied to the original program. For the remaining 3545 programs that we submitted for evaluation, 1686 programs are accepted as valid solutions, they compile and pass the functional tests, showing the capabilities of \toolname to output proper syntax and to maintain semantics. Among the 1686 accepted programs, 356 programs either improved the running time or memory consumption.

\begin{mdframed}
\textbf{RQ1}: \toolname, a model 600x and 3700x smaller than GPT-3.5-Turbo and GPT-4 is able to generate significantly more optimized programs than OpenAI models. It is capable of producing, short, target optimizations, including appropriate compiler directive addition.
\end{mdframed}

\subsection{Answer to RQ2 (Optimization Focus)}
\label{sec:result:rq2}

\begin{figure}[]
\includegraphics[width=\columnwidth]{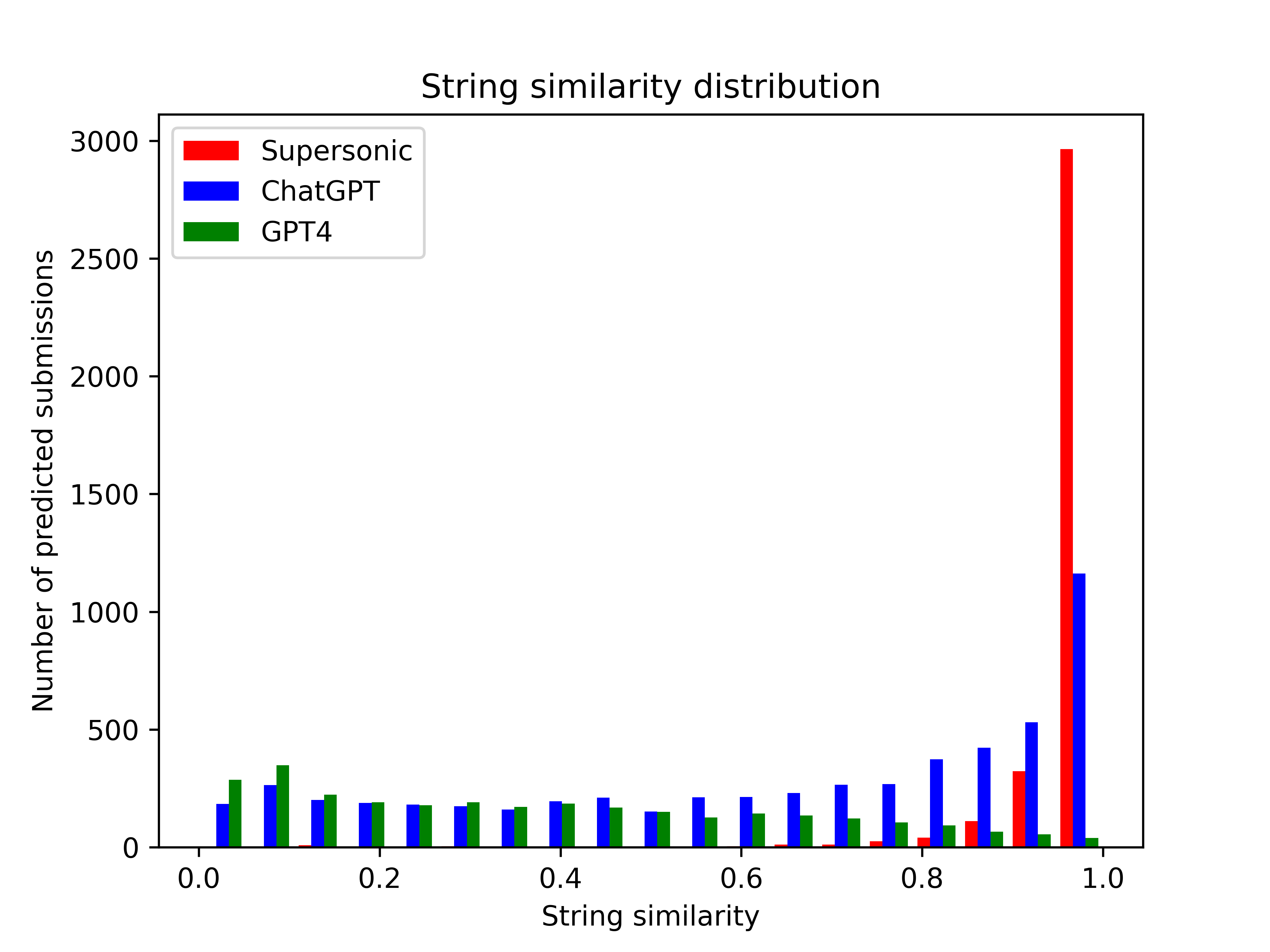}
\caption{The string similarity distribution for the similarity between the predicted submission and the original solution. Best viewed in color.}
\label{fig:rq2_time_string_similarity_distribution}
\end{figure}

\begin{figure}[]
\includegraphics[width=\columnwidth]{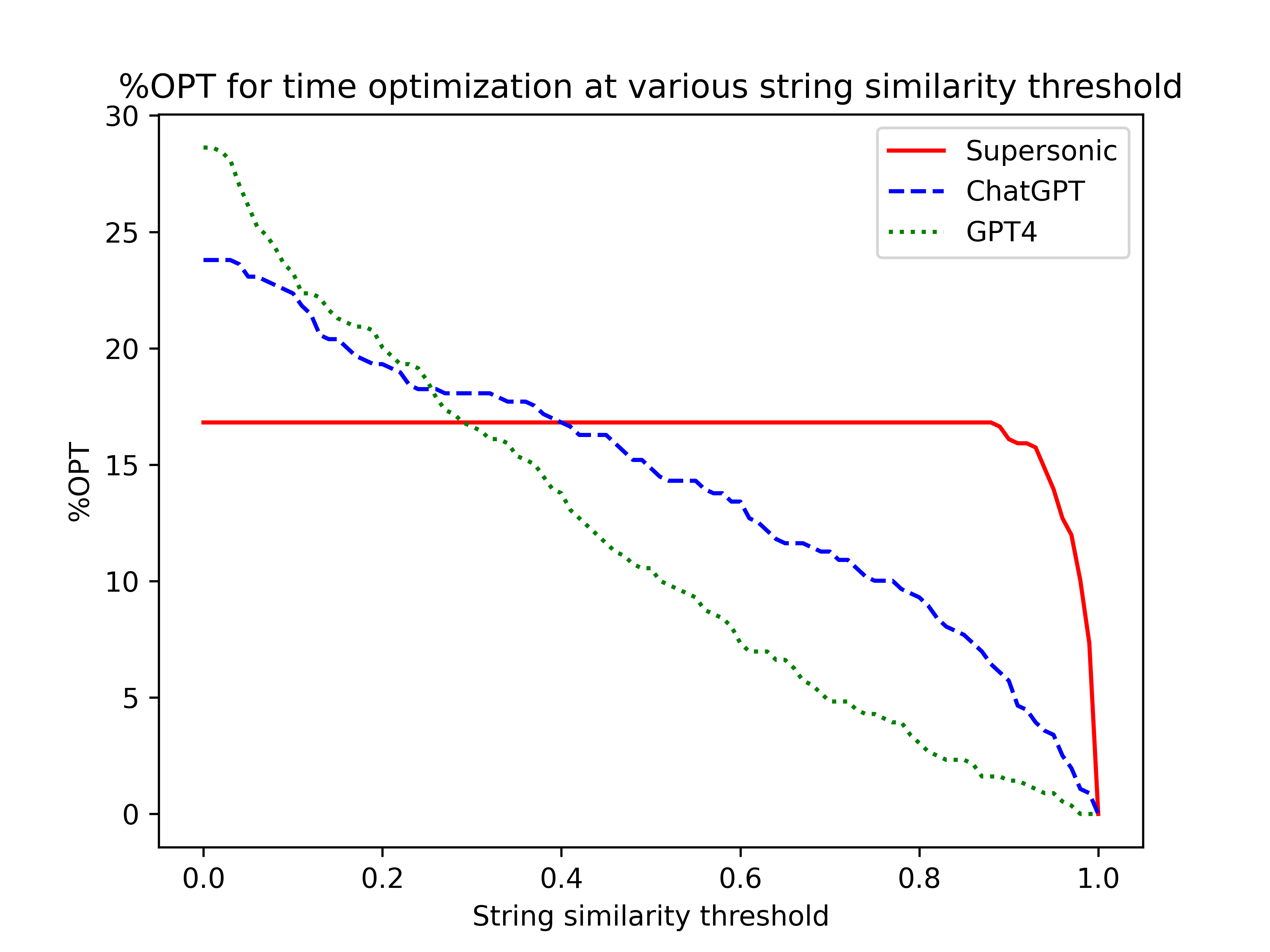}
\caption{\textsc{\%OPT} for running time optimization of \toolname, GPT-3.5-Turbo and GPT-4 if we limit the string similarity at various thresholds between the predicted submission and the original solution.}
\label{fig:rq2_string_similarity_threshold}
\end{figure}

\noindent
\begin{figure}
\begin{minipage}{\linewidth}
\begin{lstlisting}[language=diff, columns=fullflexible, frame=single, basicstyle=\footnotesize, keepspaces=true, label={lst:chatgpt_example}, caption={Unified diff for one example prediction from GPT-3.5-Turbo. Almost all source code lines are changed and the string similarity between the prediction and original program is 0.36. This cannot be considered as a targeted optimization. The original problem is the \textit{Word} problem on Codeforces\protect\footnotemark.}, captionpos=b, breaklines=true]
@@ -1,18 +1,24 @@
-#include <bits/stdc++.h>
+#include <iostream>
+#include <cstring>
 using namespace std;
+
 int main() {
-  string s;
-  cin >> s;
-  short i, t = 0;
-  for (i = 0; i < s.size(); i++) {
-    if (s[i] > 96)
-      t++;
-  }
-  i = s.size() - t;
-  if (t < i)
-    transform(s.begin(), s.end(), s.begin(), ::toupper);
-  else
-    transform(s.begin(), s.end(), s.begin(), ::tolower);
-  cout << s;
-  return 0;
-}
+    char s[1000001];
+    cin >> s;
+    int upper = 0, lower = 0;
+    for (int i = 0; i < strlen(s); i++) {
+        if (s[i] >= 'a' && s[i] <= 'z') lower++;
+        else upper++;
+    }
+    if (upper > lower) {
+        for (int i = 0; i < strlen(s); i++) {
+            s[i] = toupper(s[i]);
+        }
+    } else {
+        for (int i = 0; i < strlen(s); i++) {
+            s[i] = tolower(s[i]);
+        }
+    }
+    cout << s;
+    return 0;
+}
\end{lstlisting}
\end{minipage}
\end{figure}
\footnotetext{{\url{https://codeforces.com/problemset/problem/59/A}}}

The string similarity distribution for all generated programs is shown in \autoref{fig:rq2_time_string_similarity_distribution}. GPT-3.5-Turbo and GPT-4 generate prediction at all string similarity levels, which means they sometimes generate target optimizations and sometimes rewrite the whole program. On the contrary, most of \toolname's predictions have a high string similarity (red distribution shifted to the right), which shows \toolname generates much more precise edits to the programs under optimization. 
In addition, we see that most generated programs have a similarity higher than 0.8, which is very consistent with the training data curation, as described in \autoref{sec:tool:pre_processing}, showing the soundness of the end-to-end training and inference loop.

\autoref{fig:rq2_string_similarity_threshold} shows how the \textsc{\%OPT} metric changes with different string similarity thresholds for generated programs that optimize the running time. The y-axis is the \%OPT value. The x-axis is the string similarity threshold, meaning that we filter all generated programs with a lower string similarity score than the corresponding x value. For $x=0$, it means that we include all generated programs, and for $x=1$, all generated programs are filtered out. 
Reading the graph from right to left, we can see that \toolname is the best (red curve at the top) until the string similarity threshold is close to 0.4. This means the string similarity threshold of 0.8 in RQ1 was not a magic, unstable one. Rather, it can be changed to a large degree and \toolname would still be the best performing tool. From the same graph, we can also see that GPT-4 performs better than GPT-3.5-Turbo when the threshold is 0, but the relationship reverses when the threshold is higher. Overall, it is clear that GPT-4 is generating more full rewrites than GPT-3.5-Turbo.

To give an idea of how dissimilar two programs are when the similarity score is low, \autoref{lst:chatgpt_example} shows a GPT-3.5-Turbo generated program with similarity 0.36. It is clear that most lines are changed, including variable names. Other significant rewrites by GPT-3.5-Turbo in that example include \textit{std::transform} substituted with for-loops and \textit{char} type used instead of integer for comparison. 

\begin{mdframed}
\textbf{RQ2}: \toolname is the best performing model compared to GPT-3.5-Turbo and GPT-4 even when we lower the similarity threshold. GPT-3.5-Turbo and GPT-4 tend to optimize by fully rewriting the program which is argued to be a different task than optimization, closer to re-implementation.
To the best of our knowledge, \toolname is the best model for generating precise, targeted, and effective source code optimizations.
\end{mdframed}

\subsection{Answer to RQ3 (Ablation Study)}
\label{sec:result:rq3}

\footnotetext{\url{https://codeforces.com/problemset/problem/59/A}}

\begin{table}[]
\resizebox{\columnwidth}{!}{%
\begin{tabular}{ll|ll|ll}
\multicolumn{2}{l|}{\multirow{2}{*}{Metrics}} & \multicolumn{2}{c|}{\toolname Full} & \multicolumn{2}{c}{\toolname Diff} \\
\multicolumn{2}{l|}{}                         & Running time       & Memory       & Running time       & Memory      \\ \hline
Codeforces & \%OPT & 11.0\% (33) & 2.7\% (8) & 26.0\% (78) & 8.0\% (24) \\
           & PI    & 3.28$\times$ & 1.57$\times$ & 2.65$\times$ & 1.81$\times$ \\ \hline
AIZU       & \%OPT & 1.5\% (4) & 0.0\% (0) & 3.5\% (9) & 1.2\% (3)         \\
           & PI & 3.24$\times$ & 0.00$\times$ & 2.82$\times$ & 1.23$\times$
\end{tabular}
}
\caption{\toolname with diff-based output representation (\toolname Diff) and full program (\toolname Full) as output. Within the parentheses for \%OPT is the absolute number of optimized programs}
\label{tab:rq3:diff_vs_full}
\end{table}

For this research question, we focus on investigating the efficacy of two distinct code output representations when training our \toolname model. One version of \toolname is trained to generate an entire program as output, exemplified in \autoref{lst:improved_solution}. The second version is trained to produce a diff-based output representation, which can be seen in \autoref{lst:diff_output}. \autoref{tab:rq3:diff_vs_full} shows the comparison between the two versions on the predicted submission to the Codeforces and AIZU code competition website. 

\toolname with a diff-based output representation is clearly better than \toolname with a full program output. The diff-based output representation optimizes more programs on both Codeforces (26.0\% versus 11.0\% for running time, 8.0\% versus 2.7\% for memory consumption) and AIZU (3.5\% versus 1.5\% for running time, 1.2\% versus 0.0\% for memory consumption). Considering the presented data, it is evident that the diff-based output representation of \toolname consistently outperforms its full program counterpart in terms of generating a higher percentage of optimized programs.

This fully validates our initial hypothesis that longer output sequences are harder to learn and lower the overall performance of the model. However, the diff-based representation comes at a cost of stricter well-formedness rules for applying the diff-based output. During the breakdown of all generated programs of \toolname in RQ1, we indeed showed that a portion of diff-based output representations are malformed (2045 out of 5590). While being potentially valid optimizations, this results in fewer programs being submitted for evaluation. Overall, the advantages of the sort diff representation win over the cons, and \toolname with diff-based output representation clearly outperforms the full program output representation.

\begin{mdframed}
\textbf{RQ3}: \toolname with the original diff-based output representation doubles the performance, compared to traditional full program output representation. While the majority of works in the field of code generation with deep learning use full program or full function output, our results clearly call for more exploration of shorter, contextual output representations such as \toolname's.
\end{mdframed}

\section{Discussion}
\label{sec:discussion}

\subsection{Threats to Validity}
\label{sec:discussion:threat}

\subsubsection{Internal Threats}
\label{sec:discussion:threat:internal}

\paragraph{Data leakage.} The evaluation of GPT-3.5-Turbo and GPT-4 poses an internal threat in the form of potential data leakage that is beyond our control. The data used to train GPT-3.5-Turbo and GPT-4 is not publicly accessible, which means we lack visibility into the specific information it was exposed to during its training process. Consequently, there exists a potential risk of data leakage that GPT-3.5-Turbo and GPT-4 has already seen examples in the test set during its training process.

\paragraph{Functionally incorrect programs.} We have noticed that some original programs are functionally correct according to the original metadata, but they failed when we submitted them ourselves. For Codeforces, the main reason was that they were either because of time/memory limit, or compilation error. For AIZU, it is a mix of compile error, time/memory limit, and wrong answer. We hypothesize that it is because of the compiler version being updated throughout the year. For these programs, we still count them as failures when calculating the \%OPT value to ensure a fair comparison. 

\paragraph{Measurement uncertainty.} We also found that if we submit the same program twice to the Codeforces platform, we may get different execution times, \eg 0ms and 15ms. This is because the Codeforces platform splits the execution time as different blocks, which means 0ms and 15ms are equal. Therefore when comparing execution time for programs on Codeforces, we divided all time into blocks of 16, \ie 0 and 15 would be block 1, 16 and 31 would be block 2. Then we use the block number to calculate the speed up. Additionally, to account for noise, we also define a program to be more optimized if the time or memory improvement is at least 1.2, as mentioned in \autoref{sec:experiment:rq1}.

\subsubsection{External Threats}
\label{sec:discussion:threat:external}

A significant external threat is that CodeNet and Codeforces datasets are both from code competition websites. They do not fully represent the diversity of real-world optimization problems. The problems on competition websites are designed with code optimizations in mind, and they facilitate the comparison between submissions by having functional correctness, running time, and memory consumption as part of their result. However, real-world optimizations are likely more varied. Future work may address this threat by creating new benchmarks for code optimization. 

\subsection{Code Optimization Versus Code Synthesis}

Our extensive experience looking at predicted optimization has convinced us that the boundary between code optimization and code synthesis is hard to define. Consider a hypothetical example: an original program implements a naive algorithm to sort a list, such as bubble sort. In some cases, the distinction between optimization and synthesis is obvious, for example if the model proposes to replace the original program with a quicksort or mergesort implementation. In other cases, if the model made many changes while retaining the core concept of bubble sort, the distinction is less clear.

The distinction between code optimization and full re-implementation has received limited attention in related research. We believe that clarifying this distinction is crucial in understanding the true capabilities of neural models. In this paper, we contribute to this clarification in a fully operational way. We draw the line between optimization and re-implementation by looking at the string similarity between the original and the predicted optimized program. This is a well-formed definition: a string similarity value of 0 is pure code synthesis, and a value close to 1 is a small change to the program. 
Nonetheless, this purely syntactic solution could be improved with semantic and runtime analysis, we leave this as a topic for future exploration. 
 
\subsection{Importance of Input/Output Code Representation}

Designing the input and output representations of code for machine learning models is a critical aspect that demands careful consideration. The effectiveness of these representations can significantly impact the performance and efficiency of the resulting models.

Creating an appropriate input representation for code is akin to providing the model with the right set of tools for understanding and processing the task at hand. When designing input representations, it is essential to strike a balance between comprehensiveness and conciseness. The representation must capture the relevant information that enables the model to perform the desired task effectively. For the code optimization task, it is uncommon that we know in advance the source code location where we can optimize the program. In comparison with other code-related tasks such as the program repair task, we often have fault localization techniques that guide us toward the buggy locations. Therefore for \toolname, we decided to use the full program as input. 

Crafting a concise and precise output representation for code is akin to presenting the model's insights and solutions in a simple manner, ensuring that the generated code conveys the intended functionality without unnecessary redundancy. Redundancy in the output is repeating information that was present in the input and increases the probability of unnecessary mistakes. Such mistakes are more unforgiving than mistakes in natural languages, as programs are interpreted by compilers with strict parser rules. In \toolname, we use the diff-based output representation to minimize the redundancy and risk of mistakes and only keep two context lines to localize the change. RQ3 proves that this design choice alone improves the performance of the model by 2 times. It shows the importance of designing a proper input and output representation for machine learning for code models that are rarely discussed.

\subsection{Large versus Small Models}
It is widely accepted that larger models perform better, they are more sample efficient \cite{brown2020language}, are few-shot learners \cite{kaplan2020scaling}, and the performance increases with model size \cite{chowdhery2022palm}. Furthermore, the accuracy and the robustness of a model also increase with dataset size \cite{lei2018training}. One model that we compared against, GPT-3.5-Turbo, has at least 175B parameters and is trained on 500 billion tokens~\cite{brown2020language}. It demonstrated capabilities for many software engineering-related tasks, such as code generation~\cite{liu2023improving}, program repair~\cite{sobania2023analysis}, and code summarization~\cite{tian2023chatgpt}. In RQ2 of our study, we have also shown the good performance of GPT-3.5-Turbo on the code optimization task. However in RQ1, \toolname was able to outperform GPT-3.5-Turbo on the task of generating similar but optimized programs, despite being approximately 600x smaller (175B versus 278M) and trained with much less data. It shows the relevance of training a smaller model focusing on a single task with well-curated input and output representation.

\section{Related Work}
\label{sec:related}

\subsection{ML for Source Code Optimization}

PIE4Perf is the most related work \cite{madaan2023learning}. In this work, \citeauthor{madaan2023learning} investigates the ability of large language models to suggest performance-improving code edits. First, they extract samples of performance-improving code edits from the Codenet dataset. Then, they fine-tune the CodeGen model and they prompt Codex using few-shot prompting. They find that the system can generate performance-improving edits with speedups of more than 2.5x for over 25\% of the programs, with a model 10x smaller than Codex to match its performance. The main differences between \toolname and PIE4Perf are:
1) we study both the running time and memory optimization while PIE4Perf only looks at running time optimization. 
2) \toolname is based on an original diff-based output representation, while PIE4Perf outputs simple vanilla programs.
3) we take care of removing re-implementations in the evaluation procedure, yielding more meaningful measurements.
4) we report competitive results w.r.t. GPT-3.5-Turbo and GPT-4 with a model that is 600x and 3700x smaller.

DeepDev-PERF is a deep learning-based approach to improve software performance for C\# applications \cite{garg2022deepdev}. The model is pre-trained on both English and source code corpora and fine-tuned on the task of performance improvement for C\# applications. The evaluation shows that the model can generate the same performance improvement suggestions as the developer patches in 53\% of the cases. The authors also submitted 19 pull requests with performance optimizations, of which 11 of these are accepted. \toolname is different from DeepDev-PERF by looking at C/C++ optimizations. The evaluation is also done by submitting to code competition websites that report the running time and memory consumption, instead of using the benchmark.

\citeauthor{chen2022learning} propose a discrete variational auto-encoder to extract discrete latent variables, each representing a code-edit that increases program performance \cite{chen2022learning}. The learned discrete latent variables can be used to guide programmers toward writing more efficient source code. They show that the discrete variational auto-encoder extracts better code efficiency edits than the Transformer baseline. \toolname on the other hand does not extract code efficiency edits but directly predicts the more optimized program in an end-to-end way.

RAPGen is an approach based on the OpenAI Codex model to do zero-shot code inefficiency fixing in C\# methods~\cite{garg2023rapgen}. This is done by first collecting a dataset of performance bug fixes by keyword matching on the commit message. The dataset is used to extract identifiers that are changed in the commits to form a knowledge base. The knowledge base is used to form an instruction to describe which identifiers were added/removed/edited when giving the input method. To predict the code inefficiency fix, they build a prompt by using the buggy method as a comment, followed by the instruction of which variables to change, and ending with the buggy method's signature. The prompt is used as input to the Codex model, and it generates the fixed method. They found that it performs better than DeepDev-PERF on exact match and CodeBLEU score, without any training.

Artemis++ is a tool for syntax-based optimization of C++ using a genetic algorithm \cite{giavrimis2021genetic}. The tool automatically chooses implementations of common data structures to provide performance improvements. It uses a genetic algorithm that automatically performs source code transformations and produces an optimized version of the program. The tool is evaluated on three C++ libraries, observing improvements by up to 16.09\%, 27.90\%, and 2.74\% for CPU usage, runtime, and memory, respectively. \toolname is different in that it is trained to do more than data structure optimization, and relies on language model and seq2seq learning instead of a genetic algorithm.

LoopLearner is a tool that predicts the speedup of loop transformations \cite{mammadli2021learning}. They encode the source code as a vector and concatenate a compact encoding of transformations to the vector. The concatenated vector is then fed into CNN or RNN to predict the speedup of transformations. They found that by applying the top transformation, the program yields an average speedup of 1.29x. While LoopLearner only focuses on loops, \toolname's input the whole program and can target different source code locations.

\subsection{ML for Compiler Optimization}

\citeauthor*{ashouri2018survey} have a survey about compiler autotuning \cite{ashouri2018survey}. They summarize papers on two problems in the machine learning for compiler optimization field, 1) selecting the best optimization and 2) the phase-ordering of optimizations. \citeauthor*{wang2018machine} focus more on the different machine-learning techniques used in the compiler optimization field.

\citeauthor*{cummins2020programl} propose ProGraML, a graph-based program representation, that can be used as input to machine learning models \cite{cummins2020programl}. The representation is a union of a control flow graph, data flow graph, and call flow graph. They show that on the task of choosing a CPU or GPU to run an OpenCL kernel, it surpassed prior approaches on all metrics.

DeepTune is a tool that predicts optimization heuristics on OpenCL kernel \cite{cummins2017end}. It directly uses the source code, instead of the binary code, as the input to an LSTM model to predict heterogeneous mapping (CPU or GPU) and  OpenCL thread coarsening factor. The results show that by learning directly on source code, instead of manual features extracted from the binary code, DeepTune could match or surpass 89\% of the predictions.

\citeauthor*{narayanamurthy2016finding} use genetic algorithms to find the best set of compiler optimizations that optimized the performance while providing better error resilience \cite{narayanamurthy2016finding}. The fitness function used in the genetic algorithm measures the error resilience of a candidate optimization. The resulting program is able to achieve similar performance as -O1, -O2, and -O3 while having better error resilience.

\citeauthor*{schulte2014post} use a genetic algorithm to reduce the non-functional properties of executables such as power efficiency \cite{schulte2014post}. The fitness function used in the genetic algorithm is hardware performance counters combined into a single scalar value. They find that they can reduce on average 20\% energy usage on AMD and Intel systems. 

\citeauthor*{wang2013using} use machine learning to partition stream processing programs \cite{wang2013using}. The partition refers to dividing the program graph into clusters that are allocated to threads. They define different program features and used PCA to reduce the dimensionality to 10. Then, they use a k-means clustering algorithm to find the previous similar good partitions. They find that it achieves a 1.90x speedup over the already-tuned partition by the compiler.

\citeauthor*{churchill2017sound} propose an approach to optimize loops in the Google native Client \cite{churchill2017sound}. The approach has a bounded verifier that verifies the correctness of a loop transformation up to bound k. If it fails, it can create a counter-example that guides the search away from this transformation. Then they apply a sound verifier that uses strong loop invariants. In the evaluation, they achieve an average 25\% speedup compared to libraries shipped by Google, and the optimized program can be formally verified.

Stoke is a tool that formulates the binary superoptimization problem as a stochastic search problem \cite{schkufza2013stochastic}. The search is guided by a cost function with a correctness term and a performance term. They use Markov chain Monte Carlo to sample the binary modification. By starting with a binary that is compiled with -O0, Stoke was able to either match or outperform code produced by -O3.

\citeauthor*{bunel2016learning} improve upon Stoke by replacing the searched distribution \cite{bunel2016learning}. Stoke uses a uniform distribution to sample modifications of the program, and \citeauthor*{bunel2016learning} instead used reinforcement learning to learn the distribution from past behaviors and program semantics. By doing so, they can increase the probability of reaching the best-optimized programs and generate better programs in fewer iterations.

\citeauthor*{rotem2021profile} tackle the problem of improving the LLVM's BranchProbabilityInfo (BPI) heuristics \cite{rotem2021profile}. They first use the profile-guided optimization (PGO) workflow to compile many different programs and record their branch probabilities. These data are used to train a model of gradient-boosted trees. Once the model is trained, it can be used to predict branch probabilities on programs without PGO workflow. They find that the geometric mean speedup of the new BPI heuristic across 10 workloads is 1.016.

\citeauthor*{baghdadi2021deep} develop a deep learning-based cost model for predicting speedups of code transformations in the Tiramisu compiler \cite{baghdadi2021deep}. The training data is created by generating random programs, applying a series of code transformations, and recording the actual speedups. The deep learning model is a combination of recurrent and recursive neural networks. They find that the proposed model has only 16\% of mean absolute percentage error in predicting speedups on full programs and that it is able to find transformations that match or are better than SOTA compilers.

\citeauthor*{agakov2006using} use machine learning to speed up the process of iterative optimization on loops \cite{agakov2006using}, \ie, iterative try different compilation optimization to determine the best set of compilation optimization. They extract different features from the loop and use PCA to represent the loop with a 5-dimension vector. This vector is used to determine the nearest neighbor in the training data and extract the best compilation optimization. They found that the average speedup is 1.22x on the T1 C6713 and 1.27x on the AMD Au1500.

\subsection{Other Work on Optimization}

\citeauthor*{petke2017genetic} have a comprehensive survey about papers using genetic algorithms to improve programs \cite{petke2017genetic}. They find that genetic algorithm has been able to improve the performance of a program for a diverse set of properties, such as execution time, energy, and memory consumption.   

\citeauthor*{marco2017improving} propose an approach to estimate the memory behavior of Spark applications to allow higher server utilization and system throughput \cite{marco2017improving}. They extract 22 raw features from the application and use PCA to reduce it to 5 principal components. To predict the function that best describes the memory usage, they use K nearest neighbor to find the most similar cluster. The evaluation shows that they achieve a 1.28x improvement on system throughput and 1.68x on turnaround time.

\citeauthor*{chen2018learning} introduce a framework AutoTVM to optimize tensor operators that are used in deep learning models \cite{chen2018learning}. They use simulated annealing with a statistical cost function trained from past historical data to find more optimized low-level code. Experiments on existing frameworks show that it yields 1.2x to 3.8x performance improvement.

\citeauthor*{ahn2020chameleon} propose Chameleon that improves upon AutoTVM by using reinforcement learning to optimize tensor operators \cite{ahn2020chameleon}. Instead of using simulated annealing that relies on random walks, they use adaptive exploration by leveraging reinforcement learning. The result is that they achieve a 4.45x speed up in optimization time over AutoTVM.

\section{Conclusion} 
In this paper, we have proposed \toolname, a tool that can generate source code optimizations for C/C++ programs with targeted changes. \toolname features an original diff-based output representation that is similar to a software patch. Our experiments clearly show that \toolname outperforms GPT-3.5-Turbo and GPT-4 on the task of source code optimization. We believe that there is a large research avenue in the area of using small but specific models targeting a given software engineering task instead of using larger general models.

\newpage
\printbibliography

\newpage

\end{document}